\documentclass[twocolumn]{article}
\usepackage{graphicx}
\usepackage{amsmath}
\usepackage{amssymb}
\usepackage{ulem}
\usepackage[latin1]{inputenc}
\usepackage{pstricks}
\usepackage{pst-node}
\usepackage{pst-plot}
\usepackage{pst-grad}
\SpecialCoor

\title{Dispersive interaction between an atom and a conducting sphere}
\author{M.M.Taddei$^1$, T.N.C.Mendes$^2$, C.Farina$^1$\\
{\small $^1$Instituto de F\'isica, Universidade Federal do Rio de Janeiro, Cidade Universit\'aria,}\\
{\small Ilha do Fund\~ao, Caixa Postal 68528, 21945-970, Rio de Janeiro, Brazil }\\
{\small $^2$ Escola de Ciências e Tecnologia, Universidade Federal do Rio Grande de Norte, Natal, Brazil}
                }
\date{}
\begin{document}

\twocolumn[
\begin{@twocolumnfalse}
\maketitle
\begin{abstract}
We calculate the van der Waals dispersive
interaction between a neutral but polarizable
atom and a perfectly conducting isolated sphere in
the nonretarded case. We make use of two separate
models, one being the semiclassical fluctuating-dipoles
 method, the other using ordinary quantum mechanics.
\end{abstract}
\end{@twocolumnfalse}
]

\section{Introduction}


The existence of attractive intermolecular forces between molecules
of whatever kind has been long assumed due to the possibility of
every gas to liquefy. The first quantitative (although indirect)
characterization of these forces was done by J.D.
van der Waals in his 1873 thesis \cite{vdW} in the equation of state
for real gases, which can be written as
\begin{equation}
    \left(P+\frac{a}{V^2}\right) (V-b) = nRT \ ,
    \label{eqvdW}
\end{equation}
in which $P$, $V$, $T$ and $n$ are the pressure, volume,
temperature, and the number of mols of the gas under consideration,
respectively, and $R$ is the universal gas constant. The parameters
$a$ and $b$ -- the van der Waals constants, which vary from one gas
to another -- can be evaluated through fitting this equation with
experimental data. While the parameter $b$ relates to the finite
molecule volume, and to the fact that an individual molecule cannot
access the entire volume $V$ of the gas, the term $(a/V^2)$ relates
to an attractive intermolecular force. Such forces received the
general name of van der Waals forces. It should be stated that when
one mentions intermolecular forces as such, one assumes that the
separation between the molecules in question (or atoms, in the case
of monoatomic gases) is large enough as to exclude the overlapping
of electronic orbitals. These forces are usually distinguished in
three types, to wit: orientation, induction and dispersion van der
Waals forces, which we briefly discuss -- more detailed discussions
can be found in references
\cite{Langbein},\cite{Margenau},\cite{Milonni}.

Orientation forces occur between two polar molecules, \i.e. two
molecules possessing permanent electric dipoles, {\it e.g.} water
molecules. These forces were first computed by W.H.Keesom, in the
early twenties, considering the thermal average of the interaction
energy of two randomly-oriented electric dipoles ${\bf d}_1$ and
${\bf d}_2$, namely, $(-2/3)(d_1^2d_2^2/4\pi\epsilon_0r^6)(1/k_BT)$
for $k_BT\gg d_1d_2/4\pi\epsilon_0r^3$, where $r$ is the distance
between the dipoles (molecules) and $k_B$ is the Boltzmann constant.
Although the amount of possible attractive orientations equals the
amount of repulsive ones, once we take into account that attraction
setups correspond to smaller energies and that the Boltzmann weight
is $e^{-\mathcal E/k_BT}$ (that is, it diminishes with increasing
energy $\mathcal E$), it can be easily understood why orientation
forces are attractive. We also note that orientation forces decrease
with increasing temperature, which is natural since higher
temperatures turn repulsive orientations as accessible as attractive
ones.

It was recognized by P.Debye and others that there ought to exist an
interaction between a polar molecule and an apolar, albeit
polarizable, one, once the polar molecule induces a dipole in the
other one, giving rise to a dipole-dipole attraction force,
responsible for the induction van der Waals forces. (In fact, even a
quadrupole or higher permanent multipole can induce such a dipole
and give rise to induction forces.) The attractive character of
these forces follow from the fact that an induced dipole is parallel
to the inducing field (in the case of non-isotropically polarizable
molecules this is at least approximately true) and that the
interaction energy between an electric field and a parallel (or
close to parallel) dipole is always negative. This correlation leads
to a nonvanishing force at increasing temperatures. Moreover, we can
evaluate the dependence of the force on the distance $r$ between the
molecules recalling that the magnitude of the field ${\bf E}_1$
generated by the permanent dipole ${\bf d}_1$ is proportional to
$d_1/r^3$, and that the energy of the interaction between this field
and the second (induced) dipole (${\bf d}_2$) is of the form
\begin{equation}
    U(r) = -{\bf d}_2\cdot {\bf E}_1(r) = -\alpha E_1^2(r) \propto -\dfrac{\alpha d_1^2}{r^6} \ ,
    \label{inducao}
\end{equation}
where $\alpha$ is the molecular polarizability (assumed linear, for
simplicity). Since $U(r)$ is proportional to ($-1/r^6$), the force
has a ($-1/r^7$) behavior.

It just so happens that the correction term $(a/V^2)$ can enhance
the ideal gas approximation for each and every gas known in nature,
including gases constituted of apolar molecules (or atoms), like the
noble gases. This leads to the conclusion that there should also
exist intermolecular forces between pairs of apolar molecules.
Whereas the aforementioned intermolecular forces involving at least
one polar molecule are classically conceivable, this third kind of
van der Waals forces, the dispersion ones, occurring between two
apolar, albeit polarizable, molecules, can only be fully understood
within the framework of quantum mechanics. It was only in 1930 --
after the development of quantum mechanics -- that R.Eisenschitz and
F.London \cite{EisenLondon} demonstrated for the first time how such
a force can appear, by performing a second order perturbation theory
on a quantum system composed of two atoms. Their result can be cast in the form of the following potential
\begin{equation}
U(r)=-\dfrac{3\hbar\omega_{m0}\alpha^2}{(4\pi\epsilon_0)^24r^6} \ ,
\label{London}
\end{equation}
where $\omega_{m0}$ is the dominant transition frequency. An
important result of their work is that these forces depend on the
polarizabilities $\alpha$ of the atoms in question, which are
related to the refractive index, and consequently to the
electromagnetic dispersion in a medium composed of such atoms. The
proportionality constant in this power law could then be evaluated
by the two authors from fitted parameters of optical dispersion
measurements. This evaluation, further developed in a paper of the
same year by F.London alone \cite{London}, motivated London's
coining such forces as {\it dispersive} forces in the latter
article. This explanation successfully overturned the attempts to
base interatomic forces between apolar molecules on
permanent-quadrupole interaction, since it best fits experimental
data as the van der Waals constant $a$ itself. In London's words
\cite{London} (translated by the authors themselves):
\begin{quote}
{\it Since the van der Waals attraction, according to the previously
accepted picture, is proportional to the square quadrupole moment,
using the wave-mechanical model [for $H_2$] one obtains, in what
should be equalities, only 1/9 (according to Keesom), 1/67 or 1/206
(according to Debye) of the actual value of the constant $a$ of the
van der Waals equation.}
\end{quote}
Although the important calculation of the $(-1/r^6)$ power law is
performed in the first article, being only mentioned in the second
one, the latter is far more often cited than the former, and these
forces are also called ``London forces''.

Dispersive forces are, then, the electromagnetic forces that occur
between atoms or molecules possessing no permanent electric or
magnetic multipole whatsoever, and are due to quantum fluctuations
on the atomic charge and current distributuons. They occur not only
between two atoms, but also between macroscopic bodies, as shown for
the first time in 1932 by Lennard-Jones \cite{Lennard-Jones}, who
calculated such interaction between a polarizable atom and a
perfectly conducting plane wall. Dispersive forces can be further
divided into two kinds: nonretarded and retarded. London's work
refers exclusively to nonretarded forces, which result when one
considers light speed to be infinite, and the interaction
instantaneous. Retarded interactions, first calculated by Casimir
and Polder in 1948 \cite{CP}, take into account the finiteness of
interaction propagation speed, and in this case the dipole field of
a first molecule will only reach a second one after a time interval
of $r/c$, and the reaction field of the second molecule at the first
one will be delayed in $2r/c$. Such delay decreases the correlation
between the fluctuating dipoles, what causes the retarded force to
drop more rapidly with distance than the nonretarded one. In an
atomic system, a characteristic time is given by the inverse of a
dominant transition frequency $\omega_{mn}$, and a distance $r$ is
said to be long (or, equivalently, retardation effects become
relevant)
when $r/c \gtrsim 1/\omega_{mn}$. It should be clear that
nonretarded forces are a good approximation when the molecule
separation is small, which is the regime of validity of the London
forces.

Although Eisenschitz's and London's results were obtained by the use
of perturbative quantum mechanics, it is possible in the
short-distance limit to estimate such forces with a much simpler
method, known to have produced good results in calculations of this
kind -- as the atom-atom and the atom-wall van der Waals interactions --
that goes by the name of fluctuating-dipoles method. This method can
be found, for instance, in refs.\cite{Langbein},\cite{Milonni} and
has also been shown to be useful in enabling, with few effort and
requiring less background on quantum mechanics, various discussions
on dispersive forces, such as nonaddivity \cite{FarinaNA} or the
nonretarded force between an electrically polarizable atom and a
magnetically polarizable one \cite{Farinaalfabeta}.

Our interest in this article lies on nonretarded van der Waals
forces, and, more specifically, on the force between an atom and a
macroscopic body. We wish to further develop the calculations of
such forces by approaching a problem with curved geometry, to wit,
the nonretarded (``London'') force between an atom and a perfectly
conducting isolated sphere. We will, in fact, approach this problem
in two separate, independent ways. We first make use of the
fluctuating-dipoles method and secondly perform this calculation in
a way closer to the Eisenschitz and London approach or the
Lennard-Jones approach, making use of ordinary quantum mechanics.
This naturally leads to a more reliable result than the first, and
at the same time serves as correctness test for the
fluctuating-dipoles method.


Dispersive forces involving macroscopic bodies is of undeniable
importance for direct experimental verification, as seen in
\cite{RaskinKusch},\cite{wedge},\cite{Landragin} (check also
\cite{Dalibard} and references therein). An aspect of van der Waals
forces related to the interaction with macroscopic bodies is its
nonadditivity, which leads to the fact that one cannot, in
principle, obtain the correct van der Waals dispersion force in
macroscopic cases by simply performing pairwise integration of the
power law found for the atom-atom case. The reader interested in
this feature of dispersive forces should consult
\cite{Langbein},\cite{Margenau},\cite{Milonni}.

The modern quantum field theory explanation for such forces relies
on the fact that there is, even in sourceless vacuum, a residual
electromagnetic field whose vacuum expectation values $\langle {\bf
E}({\bf r},t)\rangle$ and $\langle {\bf B}({\bf r},t)\rangle$ are
zero, but whose fluctuations $\langle{\bf E}^2\rangle$, $\langle{\bf
B}^2\rangle$ do not amount to zero. This vacuum field can induce an
instantaneous dipole in one polarizable atom (or molecule), and the
field of this induced dipole, together with the vacuum field,
induces an instantaneous dipole in the second atom (molecule). It
can thus be said that the vacuum field induces fluctuating dipoles
in both atoms and the van der Waals dispersive interaction energy
corresponds to the energy of these two correlated zero-mean dipoles.
A more detailed analysis of dispersion forces and quantization of
the electromagnetic field can be found in Milonni's book
\cite{Milonni}. For a  pedagogical review of dispersive forces see,
for instance,  B.Holstein's paper \cite{Holstein}.


As a warm-up and to establish basic concepts and notation, the next
section is dedicated to reobtain the interaction between a
polarizable atom and a conducting plane wall using the
fluctuating-dipoles method. We then proceed in the following section
to the interaction between an atom and a perfectly conducting
isolated sphere, first by the fluctuating-dipoles method, then using
ordinary quantum mechanics. We end that section commenting the
obtained results. In a last section we make our final remarks.

\section{Calculation of the atom-wall London force}
It can be shown quantum-mechanically that the multipole of the atom
that contributes the most for this kind of interaction is the dipole
(see \cite{Bransden} for a demonstration in the atom-atom
case), thus motivating our picture of the atom as a dipole.
Furthermore, we know that this dipole is not permanent, but
fluctuates with a zero-mean value. The fluctuating-dipoles method
models the atom as constituted by a fixed nucleus and by an electron
of charge $(-e)$ and mass $m$. The binding force between them is
taken to be classical-harmonic, thus leading to a
harmonic-oscillating dipole ${\bf d}(t)$ of frequency $\omega_0$. We
take, for simplicity, the oscillation direction to be fixed, albeit
arbitrary, and so we have
\begin{equation}
    {\bf d}(t)=-e \ x(t) \ \hat{\bf x} \ ,
    \label{dipole}
\end{equation}
where $x(t)$ is position of the electron relative to the nucleus and
$\hat {\bf x}$ is the (fixed) unitary vector in the direction of
oscillation. Since the atomic polarizability is defined by the
expression ${\bf d}=\alpha{\bf E}$, one can calculate the static
atomic polarizability predicted by this model equaling the (static)
force exerted by an external electric field ${\bf E}$ to the
harmonic binding force:
\begin{eqnarray}
 -e E {\bf \hat x}=\dfrac{m\omega_0^2}{-e}\underbrace{(-ex{\bf \hat x})}_{\bf d} \ \Rightarrow \ \alpha= \dfrac{e^2}{m\omega_0^2} \ .
    \label{polariz}
\end{eqnarray}

\begin{figure}[ht]
\centering
\includegraphics[width=\columnwidth]{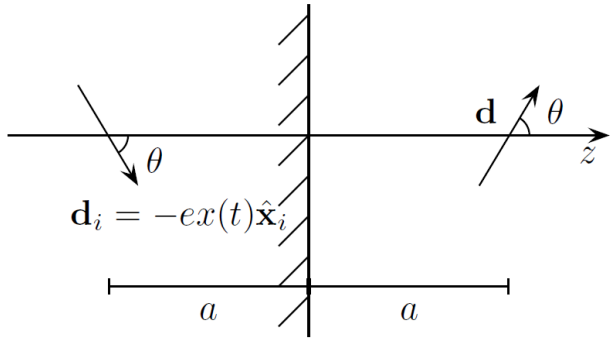}
\caption{Image configuration for a real dipole ${\bf d}$ in front of a conducting wall. The magnitudes of ${\bf d}$ and ${\bf d}_i$ are the same. Both dipoles (and the entire figure above) lie on the same plane.}
\label{fig:wall}
\end{figure}

To calculate the interaction of a dipole with a perfectly
conducting plane wall, we need to make use of the image method. The
image produced when a real dipole ${\bf d}$ stands before a
conducting plane is a dipole ${\bf d}_i$ of the same magnitude of
the real one, and whose direction is described in
fig.\ref{fig:wall}.

We now proceed to calculate the equations of motion of the electron
of the polarizable atom under the influence of the electric field
generated by the image dipole. Since the real dipole oscillates, so
does the image one, and we need, then, the field created by an
oscillating dipole, which is
\begin{eqnarray}
\nonumber {\bf E}({\bf r},t) =  \dfrac{ \left[ 3({\bf d}(t^*)\!\cdot\!{\bf\hat r}){\bf\hat r} - {\bf d}(t^*) \right] }{(4\pi\epsilon_0)r^3} \ + \\
 + \dfrac{ \left[ 3(\dot{\bf d}(t^*)\!\cdot\!{\bf\hat r}){\bf\hat r} - \dot{\bf d}(t^*) \right] }{(4\pi\epsilon_0)cr^2}+ \dfrac{ \left[ (\ddot{\bf d}(t^*)\!\cdot\!{\bf\hat r}){\bf\hat r} - \ddot{\bf d}(t^*) \right] }{(4\pi\epsilon_0)c^2r} \ ,
\label{oscillfield}
\end{eqnarray}
where ${\bf r}$ is the vector from the oscillating dipole to the
point where the field is being evaluated and
$\mbox{$t^*:=t-\frac{|{\bf r}|}{c}$}$ is the retarded time.
Fortunately, we are only interested in the small-distance
nonretarded regime, which means that we can replace the retarded
time by the time $t$ and neglect the two last terms of the rhs of
eq.(\ref{oscillfield}). The equation of motion for the electron of
the real atom, once we project the forces acting on it to the
direction of allowed motion (that is, the direction $\hat{\bf x}$ of
the dipole), becomes
\begin{equation}
\ddot x(t)+\omega_0^2x(t) = \dfrac{-e}{m4\pi\epsilon_0} \Big[ \dfrac{3({\bf d}_i\cdot \hat {\bf z})\hat {\bf z} - {\bf d}_i}{(2a)^3} \Big] \cdot \hat{\bf x} \ ,
\label{eqwallbegin}
\end{equation}
where $\hat{\bf z}$ and $a$ are defined in fig.\ref{fig:wall}. Using
the expression of ${\bf d}_i$ from fig.\ref{fig:wall}, and writing
the scalar products as a function of the angle $\theta$, we have
\begin{equation}
\ddot x(t)+\omega_0^2x(t) = \dfrac{e^2x(t)}{m4\pi\epsilon_0}\dfrac{1 + \cos^2\theta}{8a^3}
\label{eqwalltwo}
\end{equation}
But this is a simple harmonic oscillator equation, whose frequency is
\begin{eqnarray}
\omega \!\!\!&\!\!=\!\!&\!\!\! \omega_0\sqrt{1-\dfrac{e^2(1+\cos^2\theta)}{4\pi\epsilon_0m8a^3\omega_0^2}} \\
       \!\!\!&\!\!=\!\!&\!\!\! \omega_0 \left\{ 1 - \dfrac{e^2(1+\cos^2\theta)}{4\pi\epsilon_0m16a^3\omega_0^2}
 + \mathcal O
 \left(\dfrac{e^2/m\omega_0^2}{4\pi\epsilon_0a^3}\right)^2\right\}\;\;\;\;
\label{omegaalterado}
\end{eqnarray}

We now, following the chosen method, quantize the system merely
turning  classical harmonic oscillators into quantum ones of same
frequency. When the external fields alter the frequency of the
oscillator, we can quantum-mechanically say that its energy was
altered too. Now, the essence of the method lies on identifying the
zero-point energy variation as a potential energy, \i.e.,
$U:=\hbar(\omega-\omega_0)/2$. This means to compute the difference
in energy between our system as it is and the corresponding existing
system if there were no electric field and interpret that difference
as an interaction potential. Assuming the atom to be isotropic, we
replace $cos^2\theta$ by its spatial average of $1/3$. We thus find
as a leading term
\begin{equation}
U(a):=\dfrac{\hbar(\omega-\omega_0)}{2}=-\dfrac{\hbar\omega_0\alpha}{(4\pi\epsilon_0)24a^3} \ ,
\label{paredenretsemclass}
\end{equation}
and this is the atom-wall dispersive potential found by this method.
The frequency $\omega_0$ artificially introduced before is
identified with a dominant transition frequency of the atom.

Different textbooks on basic quantum mechanics present the
calculation of the atom-wall van der Waals' potential using ordinary
quantum mechanics, as for instance, the one by Cohen-Tannoudji {\it
et al.} \cite{Cohen}. One can in this fashion reobtain the
Lennard-Jones result of 1932 \cite{Lennard-Jones} for the
short-distance dispersive interaction between a polarizable
ground-state atom and a perfectly conducting wall, which can be
written as
\begin{equation}
    U(a)= - \dfrac{\langle0|d_x^2+d_y^2+2d_z^2|0\rangle}{(4\pi\epsilon_0)16a^3} \ ,
    \label{Lennard-Jones}
\end{equation}
where $d_x$,$d_y$,$d_z$ are the dipole component operators and $|0\rangle$ the ground state. If we
assume a dominant transition, the result can be cast into an
expression which equals the triple of the end result of
eq.(\ref{paredenretsemclass}). This similarity is typical for the
fluctuating-dipoles method: it yields a result whose disagreement
with quantum-mechanical results is only a constant factor. All
dependences on parameters as $a$, $\alpha$, $\omega_{k0}$ are
correctly displayed by this semiclassical method.

\section{Calculation of the atom-sphere London force}

We now calculate the van der Waals dispersive interaction between a
perfectly conducting isolated sphere and a polarizable atom. We first
reuse the fluctuating-dipoles method in this more involved geometry,
then proceed to a quantum-mechanical approach.  

\subsection{Calculation by the fluctuating-dipoles method}

\begin{figure}[ht]
\includegraphics[width=\columnwidth]{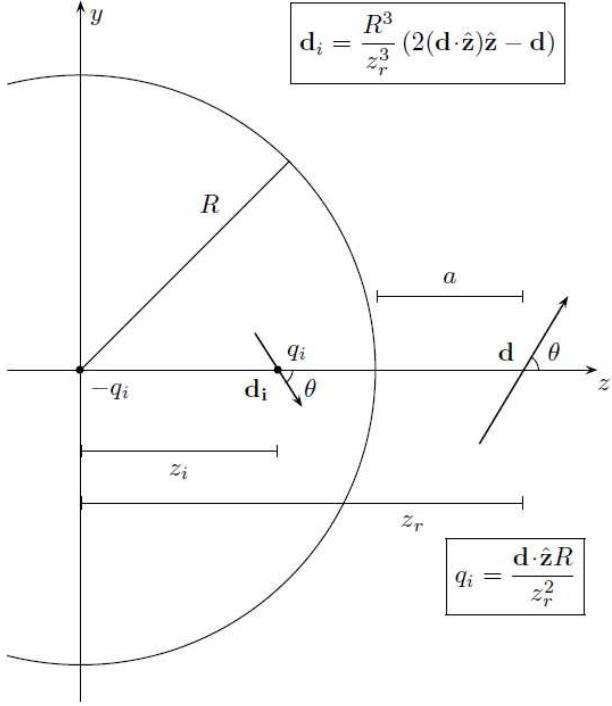}
\caption{Image configuration for a real dipole ${\bf d}$ before a conducting sphere. The values of the image dipole and image charges are indicated. $R$ is the radius of the sphere, $a$, the sphere-atom separation, $z_r:=R+a$ and $z_i:=R^2/z_r$}
\label{fig:images}
\end{figure}

All previous considerations about the atom made in the last section
apply, especially its polarizability and the interpretation of its
frequency $\omega_0$ as a dominant transition frequency. We need to
study the reaction of the sphere to the presence of a dipole, in
other words, the images produced in this situation. This classical
problem has a more complex geometry, but has been solved by
F.C.Santos and A.C.Tort in \cite{esfera}, and the solution is
described in fig.\ref{fig:images}. We define $z_r:=R+a$, where $R$
is the radius of the sphere and $a$ the minimum atom-sphere
separation. There is one image dipole ${\bf d}_i$ at radius
$z_i:=R^2/z_r$, a first image point charge $q_i$ at the same
position and a second image point charge $(-q_i)$ at the center of
the sphere. The magnitude of these images is signaled in
fig.\ref{fig:images}. Minding that, as before, forces due to the
fields of the images must be projected onto the allowed direction,
we find the following equations of motion for the electron of the
polarizable atom
\begin{eqnarray}
    \ddot x(t) + \omega_0^2 x(t)=\dfrac{-e}{m4\pi\epsilon_0}  \Big[  q_i\hat {\bf z} \left(\dfrac{1}{(z_r-z_i)^2}-\right.  \left.\dfrac{1}{z_r^2}\right) \nonumber  \\ 
+ \ \dfrac{3({\bf d}_i\cdot \hat {\bf z})\hat {\bf z} - {\bf d}_i}{(z_r-z_i)^3} \Big] \cdot \hat{\bf x} & .
    \label{eqsmotion}
\end{eqnarray}
With the values of the images to be seen in fig.\ref{fig:images},
expressing the scalar products as function of the angle $\theta$ and
using eq.(\ref{dipole}) to extract the factor $x(t)$ out of each
term, we get
\begin{eqnarray}
\ddot x(t) + \omega_0^2 x(t)=\dfrac{e^2x(t)}{m4\pi\epsilon_0} \!\!\! & \!\!\! \Big[  \dfrac{R\cos^2\theta}{z_r^2}\left(\dfrac{1}{(z_r-z_i)^2}-\dfrac{1}{z_r^2}\right)  \nonumber\\
 &\!\! + \ \dfrac{R^3}{z_r^3}\dfrac{(1+\cos^2\theta)}{(z_r-z_i)^3} \Big] \ .
\label{eqsmotsimp}
\end{eqnarray}
This is again a simple harmonic oscillator equation, whose original
frequency $\omega_0$ was altered to
\begin{eqnarray}
\omega=\omega_0 \left\{1-\dfrac{e^2/m\omega_0^2}{4\pi\epsilon_0} \right. \!\!\!&\!\!\! \Big[\dfrac{R\cos^2\theta}{z_r^2} \!\! \left(\dfrac{1}{(z_r\!-\!z_i)^2}-\dfrac{1}{z_r^2}\right) \nonumber \\ \!\!\!&\!\!\! \left. + \  \dfrac{R^3}{z_r^3}\dfrac{(1\!+\!\cos^2\theta)}{(z_r-z_i)^3} \Big] \right\}^{1/2} .
    \label{omega}
\end{eqnarray}
Replacing $\alpha= e^2/(m\omega_0^2)$ and expanding in a Taylor series
\begin{eqnarray}
    \omega=\omega_0  \left\{1   -\dfrac{\alpha}{2(4\pi\epsilon_0)} \right. \!\!\!\!&\!\!\!\! \Big[\dfrac{R\cos^2\theta}{z_r^2} \! \left(\dfrac{1}{(z_r\!-\!z_i)^2}-\dfrac{1}{z_r^2}\right) \nonumber \\  + \dfrac{R^3}{z_r^3}\dfrac{(1\!+\!\cos^2\theta)}{(z_r-z_i)^3} \Big] \!\!&\!\! \left. + \
\mathcal O \left( \dfrac{\alpha}{4\pi\epsilon_0}\xi\right)^2 \right\} \ ,
    \label{omegaseries}
\end{eqnarray}
$\xi$ being a generic form to refer to terms of $\mathcal O \left(R/z_r^4\right)$ as well as terms of  $\mathcal O(a^{-3})$, the latter occurring only when $a\ll R$. We will assume from now on the value of $(\alpha\xi/4\pi\epsilon_0)$ to be smaller than unity, and so will our calculations have this validity constraint.

Once again we turn classical harmonic oscillators into quantum ones, replace $cos^2\theta$ by $1/3$ and take the zero-point energy difference $\hbar(\omega-\omega_0)/2$ as a potential. The leading term amounts to
\begin{eqnarray}
    U=-\frac{\hbar\omega_0\alpha}{(4\pi\epsilon_0)12}  &  \left\{\dfrac{4R^3}{z_r^3(z_r-z_i)^3} \right. + & \nonumber \\ &   + \dfrac{R}{z_r^2}\left(\dfrac{1}{(z_r-z_i)^2} - \right. & \!\!\!\! \left. \left. \dfrac{1}{z_r^2} \right) \right\} \ .
    \label{potprimit}
\end{eqnarray}
 and setting the expression in terms of the parameters $R$ and $a$:
\begin{eqnarray}
    U_R(a)=-\dfrac{\hbar\omega_0\alpha}{(4\pi\epsilon_0)12}\left[\dfrac{4R^3}{(2R+a)^3a^3} +   \right. \nonumber\\
                                                    \left. + \dfrac{R}{(2R+a)^2a^2} - \dfrac{R}{(R+a)^4} \right] \ .
    \label{potdipflut}
\end{eqnarray}
The first term is due to the image dipole, the second one, to the
charge $q_i$  and the third, to the charge $-q_i$, located at the
center of the sphere.

\subsection{Calculation via perturbative quantum mechanics}

We now perform a more sophisticated, quantum-mechanical calculation
of the dispersive interaction between a ground-state polarizable
atom and a perfectly conducting isolated sphere. We shall work on
the Schrödinger picture, treat the fields classically, and deal with
quantized atoms, seeking a leading term in our potential by use of
perturbation theory.

We now again depict the atom as a fluctuating dipole, this time in a
different way. We shall state that the images' fields vary too
little with respect to position between the electron and the
nucleus, so that we evaluate the field at only one point. This is
the dipole approximation. However, some care must be taken when
writing the interaction hamiltonian between the dipole and the
conducting sphere. It can be shown that
the interaction energy of a generic configuration
of the classical system formed by a dipole and a conducting
sphere is not simply $-{\bf d}\cdot {\bf E}$, but $-(1/2){\bf d}\cdot
{\bf E}$ (see the Appendix for a demonstration of this fact, where
the energy of the configuration is computed as the external work to
bring the dipole from infinite).


Our quantized hamiltonian includes the atom hamiltonian $H_0$
(kinetic term plus coulombic attraction to the nucleus and repulsion
from other electrons) plus this interaction energy:
\begin{equation}
    H = H_0 -\frac{1}{2}\;{\bf d} \cdot{\bf E} \ ,
    \label{hamiltonianotodo}
\end{equation}
where ${\bf d}$ is now the atom's electric dipole operator, which
equals the electron charge ($-e$) times the electron's position
operator.  We shall consider the last term as a time-independent
perturbation to the atomic eigenfunctions.

We need to deal once more with the images created in a  sphere by
the presence of an electric dipole. The classical picture of
fig.\ref{fig:images} still holds, and the field at the position of
the atom can be split into two contributions, one due to the image
dipole
\begin{eqnarray}
\nonumber
{\bf E}_{{\bf d}_i}&=&\dfrac{3({\bf d}_i\cdot\hat{\bf z})\hat{\bf z}-{\bf d}_i}{4\pi\epsilon_0(z_r-z_i)^3} \\
\nonumber &=&\dfrac{-3({\bf d}\cdot\hat{\bf z})\hat{\bf z}+6({\bf d}\cdot\hat{\bf z})\hat{\bf z}+{\bf d} -2({\bf d}\cdot\hat{\bf z})\hat{\bf z}} {4\pi\epsilon_0(z_r-z_i)^3z_r^3/R^3} \\
&=&\dfrac{\left(({\bf d}\cdot\hat{\bf z})
\hat{\bf z}+{\bf d}\right)R^3}{4\pi\epsilon_0(z_r-z_i)^3z_r^3} \ ,
    \label{pifield}
\end{eqnarray}
where we used the relation between ${\bf d}_i$ and ${\bf d}$, shown
in fig.\ref{fig:images}, and the other due to both image charges,
\begin{eqnarray}
\nonumber
{\bf E}_{+q}+{\bf E}_{-q}\!\!&=&\dfrac{q_i \hat{\bf z}}{4\pi\epsilon_0(z_r-z_i)^2}+\dfrac{-q_i \hat{\bf z}}{4\pi\epsilon_0z_r^2}\\
\nonumber &=& \dfrac{q_i\hat{\bf z}}{4\pi\epsilon_0} \left(\dfrac{1}{(z_r-z_i)^2}-\dfrac{1}{z_r^2}\right) \\
&=& \dfrac{({\bf d}\cdot\hat{\bf z})\hat{\bf z}R}{4\pi\epsilon_0z_r^2} \left(\dfrac{1}{(z_r-z_i)^2}-\dfrac{1}{z_r^2}\right) \ .
    \label{qfield}
\end{eqnarray}
The perturbation hamiltonian $W$ becomes
\begin{eqnarray}
\nonumber   W=-\frac{1}{2}\;{\bf d} \cdot{\bf E} &\\
    \nonumber = - \dfrac{\left(({\bf d}\cdot\hat{\bf z})^2\!\!+\!\!{\bf d}^2\right)R^3}{4\pi\epsilon_02(z_r\!-\!z_i)^3z_r^3} \!\!\!&\!\!\! -  \dfrac{({\bf d}\cdot\hat{\bf z})^2R}{4\pi\epsilon_02z_r^2} \! \left(\!\dfrac{1}{(z_r\!\!-\!\!z_i)^2}\! -\!\!\dfrac{1}{z_r^2}\!\right) \\
    = - \dfrac{R^3\left(d_x^2\!\!+d_y^2\!\!+\!\!2d_z^2\right)}{4\pi\epsilon_02(z_r\!-\!z_i)^3z_r^3} \!\!&\!\! - \dfrac{Rd_z^2}{4\pi\epsilon_02z_r^2} \left(\!\!\dfrac{1}{(z_r\!\!-\!\!z_i)^2}\! -\!\!\dfrac{1}{z_r^2}\!\!\right) \ . \nonumber \\ \
    \label{perthamilt}
\end{eqnarray}

The first perturbative correction to the system energy is given by
$\langle0|W|0\rangle$ (with $|0\rangle$ referring to the atom ground
state). For an isotropic atom, $\langle 0|d_x^2|0\rangle=\langle
0|d_y^2|0\rangle=\langle 0| d_z^2|0\rangle$. The energy correction
is then
\begin{eqnarray}
\mathcal E^{(1)}= -\frac{\langle0|d_x^2|0\rangle}{4\pi\epsilon_02} \left\{ \dfrac{4R^3}{z_r^3(z_r-z_i)^3} \right. + & \nonumber\\
 + \dfrac{R}{z_r^2}\left(\frac{1}{(z_r-z_i)^2} \right. & \!\!\!\! - \left. \left. \dfrac{1}{z_r^2} \right) \right\} \ .
\label{PotEsfUm}
\end{eqnarray}
We identify this energy shift as the interaction potential between
an atom and a conducting sphere. If we also write the potential as a
function of the parameters $R$ and $a$, we get
\begin{eqnarray}
U_R(a)=-\dfrac{\langle0|d_x^2|0\rangle}{4\pi\epsilon_02}\left[\dfrac{4R^3}{(2R+a)^3a^3} \ + \right. &\nonumber\\
 + \dfrac{R}{(2R+a)^2a^2} \ - \!\!&\!\!\!\! \left. \dfrac{R}{(R+a)^4} \right]
\label{esferanret}
\end{eqnarray}
And once again the three terms are due to ${\bf d}_i$, $q_i$ and $(-q_i)$, respectively.

 Expressions like $\langle0|d_x^2|0\rangle$ can be calculated for
 a hydrogen atom, but, for the sake of comparison with the previous model,
 we shall assume the atom has a dominant transition frequency,
 say, $\omega_{m0}$. It has been shown (see \cite{Davydov}) that
 the static polarizability of such an atom is
\begin{equation}
\displaystyle \alpha(\omega=0)= \dfrac{2}{\hbar}\langle 0|d_x\sum_{k\neq0}\left(\dfrac{|k\rangle\langle k|}{\omega_{k0}}\right)d_x|0\rangle \ ,
\label{alfazeroexato}
\end{equation}
where $\omega_{k0}$ is the transition frequency between the ground
state $|0\rangle$ and any excited state $|k\rangle$. The dominant
transition assertion allows us to say that for every $k\neq m$ the
term $|\langle0|d_x|k\rangle|^2$ is negligible compared to
$|\langle0|d_x|m\rangle|^2$. We can then replace every $\omega_{k0}$
by $\omega_{m0}$, since for $k\neq m$ the contribution of each $k$
to the sum will be negligible, and this allows us to take
$\omega_{m0}$ out of the summation. In a two-level atom, this step
would be exact. In order to use the closure relation, we use that
$\langle0|d_x|0\rangle=0$ in an isotropic atom to add the term
$|0\rangle\langle0|$ to the sum, and we find
\begin{equation}
\hbar\omega_{m0}\alpha= 2\langle 0|d_x^2|0\rangle \ .
\label{alfazeroaprox}
\end{equation}
Hence, for a two-level atom,
\begin{eqnarray}
U_R(a)=-\dfrac{\hbar\omega_{m0}\alpha}{4\pi\epsilon_04}\left[\dfrac{4R^3}{(2R+a)^3a^3} \ + \right. &\nonumber\\
 + \dfrac{R}{(2R+a)^2a^2} \ - \!\!&\!\!\!\! \left. \dfrac{R}{(R+a)^4} \right]
\label{esferanretalfa}
\end{eqnarray}

\subsection{Discussion of the results}

We now proceed to describe the most important features of the
result, the potential given by eq.(\ref{esferanretalfa}) (or, more
generally, by eq.(\ref{esferanret})). The first striking consequence
of this result is that the semiclassical one, eq.(\ref{potdipflut}),
differs only by a factor $3$. A similar thing has already occurred
in section 2, and also happens in the atom-atom (nonretarded) case:
the results by the two methods only differ by a constant prefactor,
and this factor does not alter the order of magnitude of the
interaction. Our second result is clearly more reliable, though.

There is a very good way to check our result with the known
calculations on this subject. One only needs to take the limit
$R\rightarrow\infty$ keeping $a$ constant, in which the conducting
sphere would turn into a conducting plane wall. We do this directly
from the general expression of eq.(\ref{esferanret}), finding the potential
\begin{equation}
U_\infty(a)=-\dfrac{\langle 0|d_x^2|0\rangle}{4\pi\epsilon_04a^3} \ ,
\label{paredepelaesfera}
\end{equation}
which coincides with the previous known result given by
eq.(\ref{Lennard-Jones})  for the dispersive interaction potential
energy between an atom and a perfectly conducting plane wall in the
nonretarded limit.

Another interesting limit we can take is more peculiar, in the sense
that it corresponds to an interpretatively  challenging physical
situation: the limit $R\ll a$, in which the sphere would turn into a
so-called conducting point:
\begin{equation}
U_{0^+}(a)=-\dfrac{3\hbar\omega_{m0}\alpha R^3}{4\pi\epsilon_02a^6} \ .
\label{esferaRparazero}
\end{equation}
This is formally equivalent to the asymptotic behavior  as
$R\rightarrow0$, $a$ constant. Although the physical interpretation
of what a conducting point represents is rather subtle,
eq.(\ref{esferaRparazero}) can serve as a very useful approximation
for situations in which the conducting sphere is much smaller than
other distances in question. Furthermore, there is a strong
resemblance to the London atom-atom interaction. Since
$\alpha/4\pi\epsilon_0$ has the dimensions of a volume, it can be
loosely interpreted as an effective atom volume in its interaction
with photons. London's result (eq.\ref{London}) would consist of the
product of the transition energy with both effective volumes over
$r^6$. The corresponding volume of the sphere would be its real
volume, and, prefactors aside, eq.(\ref{esferaRparazero}) also
consists of the transition energy times the effective volumes of the
atom and the sphere over $a^{6}$. These results are, in that sense,
equivalent to each other.

\begin{figure}[ht]
\includegraphics[width=\columnwidth]{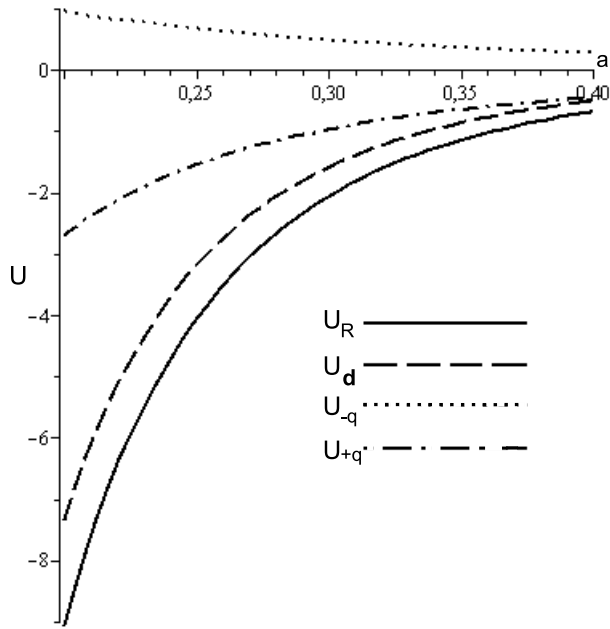}
\caption{Graph of potential between atom and sphere, eq.(\ref{esferanret}). We set  $\dfrac{\langle0|d_x^2|0\rangle}{4\pi\epsilon_02}=1$, $U_R(a)$ is in arbitrary units and $R$ and $a$ are in the same arbitrary units. We separated three contributions for the potential, $U_{\bf d}$ due to image dipole, $U_{+q}$ due to the closer image point charge and $U_{-q}$, due to the farther image charge.}
\label{graph}
\end{figure}

We lastly present a graph of the obtained potential as a function
of the distance $a$ between the atom and the sphere,
fig.\ref{graph}. Using arbitrary units, we set the prefactor
$\dfrac{\langle0|d_x^2|0\rangle}{4\pi\epsilon_02}$ equal to unity,
and $R=0.5$, where $a$ has the same units as $R$. The graph is
monotonic, and  we separate it in three terms, $U_{\bf d}$, the
contribution from the image dipole, $U_{+q}$, the contribution
from the image charge closer to the surface, and $U_{-q}$, the
contribution from the image charge at the center of the sphere. One
can clearly see that the term $U_{-q}$ contributes to a repulsive
potential, although the total interaction is always attractive. It
is intuitive, and can be demonstrated in eq.(\ref{esferanret}), that
$|U_{-q}|<|U_{+q}|$, since the charge $-q_i$ is farther from the
dipole. As the value of $R$ increases, the total interaction for
each $a$ increases in modulus, although the images grow farther from
the atom. This happens because with increasing $R$ the graph becomes
dominated by the term $U_{\bf d}$, and the image dipole has a
$(R/z_r)^3$ dependence. The terms $U_{+q}$ and $U_{-q}$ actually
decrease (in modulus) with increasing $R$.

\section{Final remarks}

It should be stated that all three kinds of van der Waals forces
have many different practical and theoretical applications, for
instance, in condensation and crystallization, in structural and
energetic effects in colloidal chemistry or biology, in the vast
field of adhesion (including its applications in washing -- the role
of a detergent is to diminish the van der Waals forces between dirt
and tissues), in its connection to the Casimir effect, among others.

We have performed the calculation of the van der Waals nonretarded
dispersive force between a polarizable atom and a perfectly
conducting isolated sphere. It constitutes an interesting
application of the fluctuating-dipoles method and of the ordinary
quantum-mechanical van der Waals calculations in situations with
curved geometry. We performed one important accuracy check on our
result, which is the limit $R\rightarrow\infty$, in which the sphere
turns into a plane. Our final result does agree with the literature
in that limit.

The most remarkable feature of our result is the calculation  of the
force when the sphere turns into a conducting point. Besides its
possible approximative value, the ideia of a conducting point may be
useful in more complex situations, as for instance in the simulation
of defects in field theory.

Our first perspective is to find the {\it retarded} dispersive
interaction between an atom and a conducting sphere. A reader might
be attempted to consider the calculation of the long-distance,
retarded interaction by use of the fluctuating-dipoles method or of
perturbative quantum mechanics without neglecting the radiation
fields as we did after eq.(\ref{oscillfield}). However, the
retardation effects require the use of a more complete description,
one including the quantized fields. We also envision to work with
dispersive interactions with curved geometries in general, such as
rugged surfaces, and to generalize our results for materials of
limited conductivity, such as dielectrics.

\section*{Acknowledgments}
The authors wish to thank both P.A.Maia Neto and A.Tenório for the
enlightening  discussions, as well as CNPq (Brazil's National
Research Council) and Faperj (Research Support Foundation of the
State of Rio de Janeiro) for partial financial support.

\section*{Appendix}
We now calculate the energy of a classical system composed of a
dipole and an  isolated conducting sphere. We shall do this
computing the external work required to bring the dipole from
infinity into its final position using a particularly simplifying
path. Our result does not lack generality, though, since the end
configuration is arbitrary and, as we know, this work is
path-independent.

We first bring the dipole from infinity keeping it parallel to the
$y$  direction (see fig.\ref{fig:images}), or keeping
$\mbox{$\theta=\pi/2$}$. In this setup we only have one image, the
dipole, of the form ${\bf d}_i=-{\bf d}R^3/(R+a)^3$. The field
generated by this image is
\begin{equation}
{\bf E}({\bf r}')=\dfrac{3({\bf d}_i\cdot\hat{\bf r}')\hat{\bf r'} -
{\bf d}_i}{4\pi\epsilon_0r'^3} = \dfrac{R^3d[ 3(-y/r')\hat{\bf r}' +
\hat{\bf y} ]}{(R+a)^34\pi\epsilon_0r'^3} \label{campogeral}
\end{equation}
where ${\bf r}'$ is the vector from the image to the point of
evaluation  of the field, and $r'=\sqrt{(z-z_i)^2+y^2}$. We shall
need the field for $y\neq0$ to derivate the field in the next step.
The force on the dipole obeys, on our case,
\begin{eqnarray}
{\bf F}&=&({\bf d}\cdot {\bf \nabla}){\bf E}=(d\,\hat{\bf y}\cdot{\bf \nabla}){\bf E}=d\,\dfrac{\partial}{\partial y}{\bf E} \\
&=& \dfrac{R^3d^2}{(R+a)^34\pi\epsilon_0} \left[-\dfrac{3\hat{\bf
r}'}{r'^4} + y(...)\right]_{y=0} \ ,
    \label{forca}
\end{eqnarray}
where $(...)$ is a nonsingular vector quantity, and
\begin{equation}
{\bf F}=\dfrac{-3d^2\hat{\bf
z}R^3/(4\pi\epsilon_0)}{(z_r\!-\!z_i)^4(R+a)^3} =
\dfrac{-3d^2\hat{\bf
z}}{4\pi\epsilon_0}\dfrac{R^3(R+a)}{a^4(2R+a)^4}
    \label{forcafinal}
\end{equation}

The work done in this first step is $W_I=-\int{\bf F}\cdot d{\bf r}$
\begin{eqnarray}
W_I\!\!\!\!&\!\!=\!\!&\!\!\!\!\!\dfrac{3d^2R^3}{4\pi\epsilon_0}
\!\int_\infty^d\!\!\dfrac{(R+a')da'}{d'^4(2R+a')^4}
 \!=\!\dfrac{3d^2}{4\pi\epsilon_0R^3}\int_\infty^{\frac{a}{R}}\!\!\dfrac{(1+\xi)d\xi}{\xi^4(2+\xi)^4} \nonumber \\
&=&\dfrac{-3d^2 \ \ \ \ R^3}{4\pi\epsilon_0 6 a^3 (2R+a)^3} =
\dfrac{-d^2 R^3/(4\pi\epsilon_0)}{ 2 (z_r\!-\!z_i)^3 (R+a)^3} \ .
\label{work}
\end{eqnarray}
The reader challenged by the above integral can make use of analytical integration softwares, such as Maple or Mathematica, to find its surprisingly simple result.

We now proceed to rotate the dipole into its final position, that
is, from $\theta=\pi/2$ to an arbitrary $\theta$ in
fig.\ref{fig:images}, and calculate the work done by the torque on
the dipole. Besides the image dipole, image point charges $q_i$ and
$-q_i$ are now present,
\begin{equation}
q_i=\dfrac{d_zR}{z_r^2} \ \ \ ; \ \ \ {\bf d}_i=(d_z\hat{\bf z}-d_y\hat{\bf y})\dfrac{R^3}{(R\!+\!a)^3}
\label{imgapp}
\end{equation}

The field on the real dipole is
\begin{eqnarray}
{\bf E}\!\!\!\!&\!\!\!\!=\!\!\!\!&\!\!\!\!\!\dfrac{3({\bf d}_i\cdot\hat{\bf z})\hat{\bf z}-{\bf d}_i}{4\pi\epsilon_0r'^3} +
\dfrac{q_i\hat{\bf z}}{4\pi\epsilon_0}\left( \dfrac{1}{r'^2} - \dfrac{1}{z_r^2} \right) \\
&=\!\!&\!\!\! \dfrac{R^3d_y\hat{\bf y}}{(R\!+\!a)^34\pi\epsilon_0r'^3} + \dfrac{2R^3d_z\hat{\bf z}}{(R\!+\!a)^34\pi\epsilon_0r'^3} + \nonumber \\
& & + \dfrac{d_z R\hat{\bf z}}{4\pi\epsilon_0z_r^2}\left(\frac{1}{r'^2}-\frac{1}{z_r^2} \right)
\label{campotorque}
\end{eqnarray}

The torque on the dipole is of the form ${\bf d}\times{\bf E}$, and
its only  relevant component is on the x axis (out of the page). We
thus need to know $d_y E_z - d_z E_y$, which equals

\begin{equation}
\dfrac{d_yd_z}{4\pi\epsilon_0} \left\{\dfrac{R^3}{r'^3(R\!+\!a)^3} + \dfrac{R}{z_r^2}\left(\frac{1}{r'^2}-\frac{1}{z_r^2} \right)\right\}
\label{pypz}
\end{equation}

The work on this second step is
$$W_{II}=-\int({\bf d}\times{\bf E})\cdot d{\bf \theta}=-\int({\bf d}\times{\bf E})_x d\theta$$

All the terms in curly brackets on eq.(\ref{pypz}) are taken out of
the integral.  Using that $d_y=d\, \sin\theta$ and $\mbox{$d_z=d\, \cos\theta$}$, the integral to perform is

\begin{equation}
d^2\int_{\pi/2}^{\theta}\sin\theta'\cos\theta'd\theta'=d^2\dfrac{\cos^2\theta}{2}=\dfrac{d_z^2}{2}
\ , \label{integral}
\end{equation}
where in the last expression (and from now on) $d_z$ is the final dipole component.
\begin{equation}
W_{II} = \dfrac{-d_z^2}{4\pi\epsilon_02} \left\{\dfrac{R^3}{r'^3(R\!+\!a)^3} + \dfrac{R}{z_r^2}\left(\frac{1}{r'^2}-\frac{1}{z_r^2} \right)\right\}
    \label{work2}
\end{equation}

Since ${\bf E}_q \parallel \hat{\bf z}$ and $r'=z_r\!-\!z_i$, one
can recognize,  comparing to eq.(\ref{qfield}), the last term as
$\mbox{$-(1/2){\bf d}\cdot({\bf E}_{+q}+{\bf E}_{-q})$}$. Summing
the first term with $W_I$ from eq.(\ref{work}) and using that
\begin{equation}
\dfrac{R^3(d^2 +
d_z^2)}{(R+a)^3}=\dfrac{R^3(3d_z^2+d_y^2-d_z^2)}{(R+a)^3}= {\bf
d}\cdot[3({\bf d}_i\cdot\hat{\bf z})\hat{\bf z}-{\bf d}_i]
    \label{pepz}
\end{equation}
we find that
\begin{equation}
W_I+W_{II}=-\frac{1}{2}{\bf d}\cdot[{\bf E}_{{\bf d}_i} +
 {\bf E}_{+q}+{\bf E}_{-q}]=-\frac{1}{2}\,{\bf d}\cdot{\bf E}
    \label{final work}
\end{equation}
justifying eq.(\ref{hamiltonianotodo}). Our result holds for any
value of $R$,  which includes the limit $R\rightarrow\infty$, \i.e.,
the plane wall. In the case of the wall, where we only have one
image, the dipole, it is quite intuitive that there be a factor
$(1/2)$. One must only consider the fields' energy density, and the
fact this density is zero in half of the space (the wall itself) and
equal as in the case of two real dipoles in the other half.
Therefore, the energy of the dipole-wall configuration is, by
symmetry, half the energy of two appropriately correlated real
dipoles. The spheric conductor does not feature such symmetry, thus
requiring the calculation in this appendix.

\end{document}